# Interaction of surface cations of cleaved mica with water in vapor and liquid forms

Giada Franceschi,*[a] Sebastian Brandstetter,[a] Jan Balajka,[a] Igor Sokolović,[a] Jiri Pavelec,[a] Martin Setvín,[b] Michael Schmid,[a] and Ulrike Diebold*[a]

Natural minerals contain ions that become hydrated when they come into contact with water in vapor and liquid forms. Muscovite mica – a common phyllosilicate with perfect cleavage planes – is an ideal system to investigate the details of ion hydration. The cleaved mica surface is decorated by an array of $K^+$ ions that can be easily exchanged with other ions or protons when immersed in an aqueous solution. Despite the vast interest in the atomic-scale hydration processes of these $K^+$ ions, experimental data under controlled conditions have remained elusive. Here, atomically resolved non-contact atomic force microscopy (nc-AFM) is combined with X-ray photoelectron spectroscopy (XPS) to investigate the cation hydration upon dosing water vapor at 100 K in ultra-high vacuum (UHV). The cleaved surface is further exposed to ultra-clean liquid water at room temperature, which promotes ion mobility and partial ion-to-proton substitution. The results offer the first direct experimental views of the interaction of water with muscovite mica in UHV. The findings are in line with previous theoretical predictions.

## Introduction

The interaction of mineral surfaces with water rules a myriad of important natural phenomena, including dissolution and weathering,[1,2] which forms soils and removes $CO_2$ from the atmosphere;[3] adsorption of toxic ions in groundwater, relevant for environmental remediation;[4] and ice nucleation on mineral dust,[5–7] which regulates cloud formation and weather patterns. The hydration of the mineral surface ions underpins all these processes. Nowadays, there is great interest in understanding the atomic-scale details of ion hydration at mineral surfaces. Still, only a handful of experimental studies have shared direct views of molecular-level ion hydration structures under pristine (ultra-high vacuum, UHV) conditions.[8]

Muscovite mica ("mica", in the following) is a common phyllosilicate with alternating $K^+$ and aluminosilicate sheets (see Figs. 1b, c), well suited to model ion hydration at mineral surfaces. Mica cleaves easily and yields virtually step-free, atomically flat surfaces[9] ideal for scanning probe microscopies. A recent nc-AFM study performed in UHV has shown that cleaving mica along (0001) plane leaves an array of short-range-ordered $K^+$ ions at the surface (see Fig. 1a for a representative image).[10] These ions can be easily exchanged with other ions or protons upon immersion in solution without modifying the underlying surface.[11–15] Many water-centered studies have been performed as a function of these substitutions, including water adsorption,[16–20] the structure of interfacial layers in thin liquid films and bulk liquids,[14,20–27] the mobility[13] and atomic-scale arrangement[21,25,28,29] of hydrated ions, and heterogeneous ice nucleation.[30–32]

To date, there have been no direct experimental observations of the hydration of the surface $K^+$ ions in UHV. The current molecular-level understanding is based on theoretical studies and simulations.[16,17,19–21,31] The lack of experimental data is mainly because the surfaces of clean mica are locally charged and difficult to image with nc-AFM.[10,33] The situation differs in ambient or liquid environments, where the interaction with the surrounding screens the surface charges and facilitates imaging. Many groups have successfully employed AFM in liquid to resolve the hydration structures of the $K^+$ ions on mica immersed in solutions.[23,25,26,29,34]

At the same time, many theoretical and experimental works have shown that the $K^+$ ions are (at least partially) washed away upon immersion in liquid and that protons take up the former $K^+$ sites to keep the system locally charge neutral.[13,15,35,36] However, there is evidence that potassium carbonate ($K_2CO_3$) forms at the surface when mica is cleaved in air due to the reaction of the surface K ions with $CO_2$ and $H_2O$, and is dissolved upon immersion in liquid.[37] To date, it is unclear under what conditions and to what extent protons substitute the surface $K^+$ ions and whether $K_2CO_3$ formation mediates this process.

This experimental study combines constant-height, non-contact (nc) AFM with X-ray photoemission spectroscopy (XPS) to obtain views of the atomic-scale hydration structures of $K^+$ ions on mica surfaces cleaved in UHV and their behavior upon immersion in liquid water under pristine conditions. Water vapor dosed in UHV at 100 K hydrates these cations through various adsorption geometries. Exposure to ultraclean liquid water in a UHV-based environment partially removes the K$^+$ cations, substituting them with protons even in the absence of CO$_2$(g), i.e., not mediated by K$_2$CO$_3$ formation.

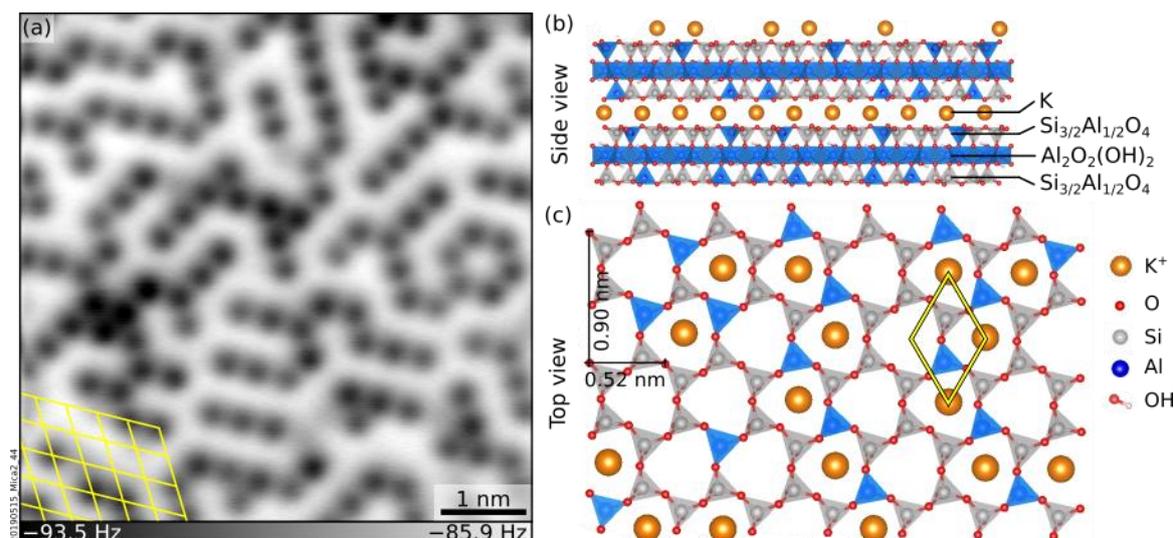

**Fig. 1. Freshly cleaved mica.** (a) Constant-height, nc-AFM image of a UHV-cleaved mica surface acquired at 4.7 K with a qPlus sensor and a metal-terminated tip. 6 × 6 nm$^2$, $A$ = 150 pm, $V_s$ = −9.3 V. K$^+$ ions at the surface appear dark. (b, c) Side and top view of cleaved mica. (b) K layers alternate with aluminosilicate layers consisting of two sheets of Si and Al tetrahedra sandwiching one sheet of Al octahedra plus OH groups. Panel (c) shows the top view of a tetrahedra sheet after cleaving. The tetrahedra are arranged in rings with quasi-hexagonal symmetry and, in the bulk, K$^+$ ions sit at the center of each ring. After cleaving, roughly 50% of the K$^+$ ions remain on each cleaved surface.

## Results and discussion

### Exposure to water in UHV

This section investigates the hydration of the K$^+$ ions with atomic precision by dosing sub-monolayer amounts of H$_2$O vapor on UHV-cleaved mica. Initially, water was dosed at room temperature (RT). However, no difference could be detected in nc-AFM between the cleaved surface and a surface dosed with 150 L water at RT (1 Langmuir = 10$^{-6}$ torr s). This suggests that water does not adsorb in UHV at RT. Lowering the sample temperature promotes water adsorption (see below). However, warming up the sample to RT after low-temperature water adsorption is not sufficient to restore the surface to its pristine state; as shown in section S1 ESI, some of the previously adsorbed water persisted.

Figure 2 presents nc-AFM images of UHV-cleaved mica surfaces after dosing sub-monolayer amounts of water at 100 K. Figures 2a, d were obtained with a nominal dose of 0.15 L, roughly corresponding to 0.03 H$_2$O/u.c. (see Experimental section). The images in Fig. 2a, d were acquired in the same area with a metal tip and a CO-functionalized AFM tip, respectively. Both images show species of different heights protruding above the flat surface. On the flat surface in between, the same meandering arrangement of K$^+$ ions of the freshly cleaved surface can be distinguished (the ion lattice here appears fainter compared to Fig. 1a because of the larger tip-sample distance; see section S2 ESI for an example of different contrast where the K$^+$ ions of the dry surface are clearly visible). The K$^+$ ions in the background are imaged dark with both tips. On the other hand, the protruding species – occupying ≈1% of the K$^+$ sites of the dry surface – appear different. With the metal tip (Fig. 2a), they appear as dark (attractive) features at all explored tip-sample distances. Approaching the tip closer to the surface than in Fig. 2a causes a strong interaction with the protruding species inducing surface rearrangements and tip changes without any resolution improvement. With the CO tip (Fig. 2d), the highest species appear as one or more bright (repulsive) round features on top of a dark (attractive) region. When the water dose is larger than 0.2 L, differently protruding species form at the surface and interact with the tip, making it challenging to retain the CO molecule at its apex. Figures 2b, c correspond to nominal doses of 0.4 L and 0.8 L and were acquired with metal tips.

As seen from Fig. 2b and the examples in Figs. 2e–g acquired at the closest possible tip-sample distance, all protruding species imaged with CO-functionalized tips share the common trait of displaying both repulsive (bright) and attractive (dark) parts. However, small differences are always present among them regarding their height, the number and orientation of bright features, and whether they appear isolated or in small aggregations. A commonly observed species (black arrow in Fig. 2e) consists of an attractive (dark) part plus two repulsive (bright) features oriented along one of the three low-index directions of mica (similar species are indicated by red and yellow circles in Fig. 2d).

The protruding species are identified as hydrated cations based on several pieces of evidence. The first indication comes from previous nc-AFM images of hydrated Na$^+$ ions on NaCl.[38] The hydrated cations were measured with a CO-tip and displayed a similar contrast as the protruding species in Fig. 2 (repulsive plus attractive part). Based on a comparison with AFM simulations, the dark and bright features were assigned to the Na$^+$ cation and H$_2$O molecules sensed by the quadrupole-like CO tip.[38] The rough calibration of the water doses fits reasonably well with this picture. At 0.15 L, ≈1% of the K$^+$ ions appear as hydrated by ≈2 water molecules. If the H$_2$O molecules only attach to K$^+$, this would correspond to ≈0.01 H$_2$O/u.c., in reasonable agreement with the calibration of 0.03 H$_2$O/u.c. from previous experiments (see above). A second indication is that DFT[17] and MD[19] calculations predict solvation forces to dominate the wetting of mica at low temperatures, i.e., H$_2$O molecules should hydrate the K$^+$ ions rather than forming H-bonded networks. Third, MD simulations[18–20] predict a large variety of hydrated species (in terms of their size and adsorption geometries of the H$_2$O molecules), consistent with the experimental observations. This variety is expected due to the heterogeneity of the cleaved mica surface. As visible in Fig. 1a and discussed in more detail in Ref. 10, the surface lacks long-range order. Different K$^+$ binding sites are available for H$_2$O molecules. Each site is characterized by a different number and arrangement of the neighboring ions and, hence, a different local electrostatic potential. This potential will determine a unique interaction with the polar H$_2$O molecule and enforce specific adsorption configurations. Fourth, it was observed that surface cations become more mobile with increasing number of hydrating molecules.[38,39] This high mobility could be due to a weaker bond of the ions to the surface after hydration, in line with previous predictions that H$_2$O molecules should lift the ions[17] (note, however, that the lifting behavior is largely neglected in the literature). The lifting is consistent with the darker appearance (stronger interaction with the tip) of the hydrated cations compared to the dry ions in the background. Ion lifting upon low-temperature water exposure is not unique to mica. Computational studies have predicted a similar pulling effect for Mg ions on MgO substrates[40] and Cl ions on NaCl.[41]

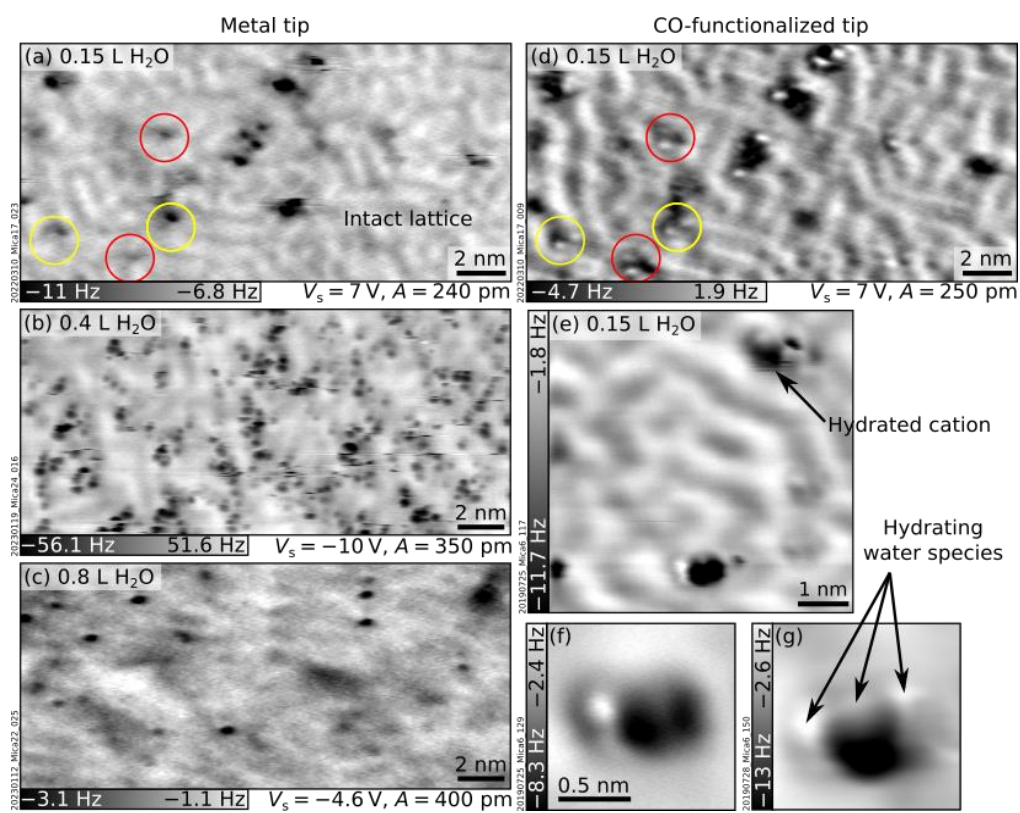

**Fig. 2. Mica after exposure to water vapor at 100 K.** AFM images acquired at $T$ = 4.7 K with metallic (left) and CO-functionalized (right) tips of UHV-cleaved mica plus sub-monolayer amounts of water vapor dosed at 100 K (doses expressed in Langmuir, L). (a, d) 20 × 9.8 nm$^2$ images of the same area after dosing 0.15 L H$_2$O. Circles highlight the most representative protruding features. (b, c) 20 × 9.3 nm$^2$ images after 0.4 L and 0.8 L H$_2$O, respectively. (e–g) Examples of hydrated cations. (e) 6 × 6 nm$^2$, (f, g) 2.5 × 2.5 nm$^2$, $A$ = 50 pm, $V_s$ = 8 V.

Finally, the XPS data in Figure 3 corroborate the overall picture. Figures 3a, b show the evolution of the K 2$p$ and O 1$s$ peaks with increasing water coverage dosed at 100 K on UHV-cleaved mica, respectively. Figures 3c, d quantify the evolution of the fitting components. The O 1$s$ peak can be fit by only two components at all coverages: the component of the dry surface (given by peaks 1 and 2 in Fig. 3a, assigned to bulk O and bulk OH) and a component 3 growing larger at increasing doses. At 50 L, where all the water should adsorb molecularly in water ice, the O 1$s$ peak can be fit with component 3 alone, which is thus assigned to molecular H$_2$O. The fact that all other O 1$s$ peaks can be fit by the components of the pristine surface and of molecular H$_2$O supports the hypothesis that water adsorption is molecular at all coverages. Note that dissociative adsorption at very small coverages cannot be completely ruled out: at 0.2 L, a reasonable fit can be obtained without component 3 (however, already at 0.4 L, component 3 is needed). The upper limit for dissociative adsorption at 0.2 L can be derived as roughly 0.1 OH/u.c..

The behavior of the K 2$p$ XPS peak is overall consistent. Its shape changes with increasing water dose, featuring a more pronounced separation of the spin-split components. As shown previously,[10] the K 2$p$ peak of the dry surface can be fit by two components assigned to different types of K ions: the fully coordinated ions in the bulk holding together the aluminosilicate layers, and the lower-coordinated ions at the cleaved surface that produce a core-level-shifted peak at higher binding energy. In Fig. 3b, the bulk and surface components are labeled as 1 and 2, respectively. Above 3 L, the K 2$p$ spectrum is successfully fit by the bulk component alone, suggesting that the K$^+$ ions are hydrated by H$_2$O molecules to reach again full coordination and a bulk-like character. Between 0.2 L and 1.6 L, an additional component (3) at a binding energy between the surface and the bulk components is needed to fit the spectra. This suggests that, at these coverages, the number of H$_2$O molecules hydrating the ions is still insufficient to complete their shell. This agrees with the notion emerging from the AFM data that the $K^+$ ions become progressively more hydrated at larger water doses, and with previous MD studies at room temperature predicting that $H_2O$ molecules preferentially adsorb close to the $K^+$ ions to complete their hydration shell.[19]

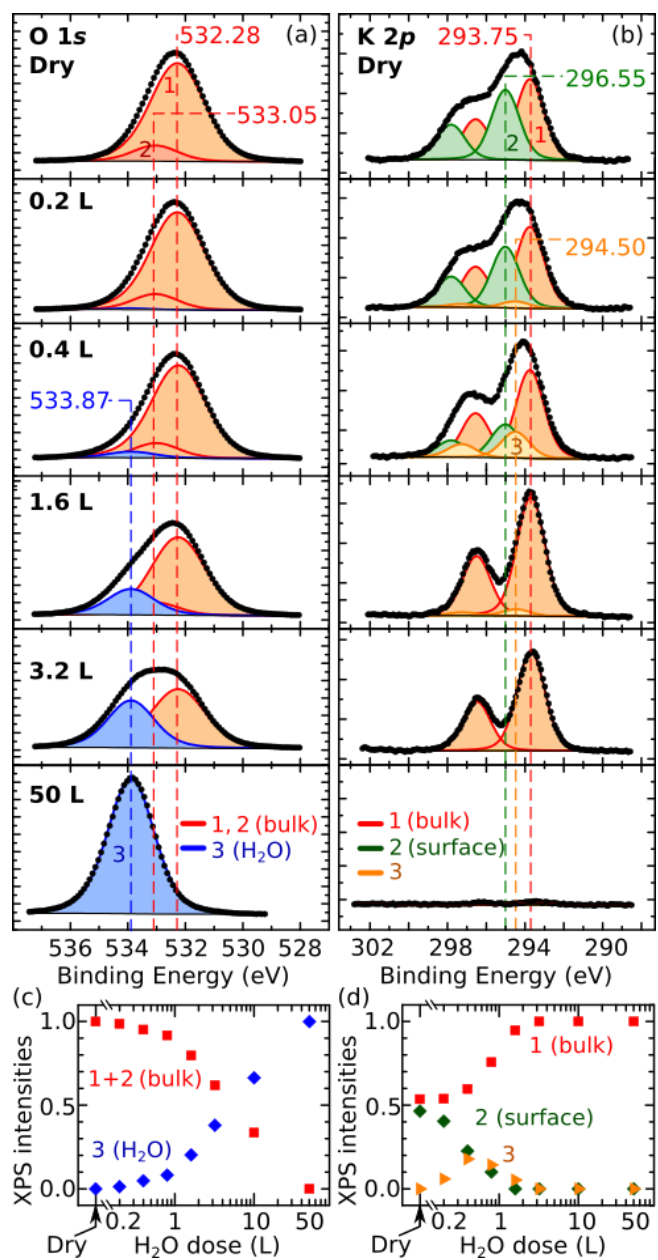

**Fig. 3. XPS after exposing mica to water vapor at 100 K.** (a, b) Experimental (dots) and fits (solid lines) of O 1s and K 2p core-level peaks of UHV-cleaved mica exposed to water vapor at 100 K (Al Kα, pass energy 20 eV, 70° grazing emission; doses expressed in Langmuir; XPS data acquired at 100 K). (c, d) Normalized intensities of the fitting components of the O 1s and K 2p core-level peaks as a function of the water dose. Error bars are smaller than the symbols size. The Experimental section provides details about the correction of the energy axes, fitting procedures, and error bars calculations.

Note that the lack of a flat, ordered, ice-like phase emerging from the AFM data is in line with previous MD simulations predicting that 2D ice structures should not form on K-mica.[19,32] Consistently, Figs. 2b, c show that increasing water doses do not produce any long-range order expected for a 2D ice layer but only the enlargement of the clusters of hydrated species, which roughly follow the short-range arrangement of the underlying $K^+$ ions. No pattern is visible in the corresponding Fourier-transformed images.

A remark about the possible role of intrinsic impurities is due. MD studies suggest that multivalent cations such as $Ca^{2+}$ and $Mg^{2+}$ should be better ice nucleators than $K^+$.[31] Hence, it cannot be excluded that the few hydrated species present at low coverages are, in fact, hydrated trace impurities. Nonetheless, the behavior of the XPS K 2p peak and the appearance of the surface in AFM at larger $H_2O$ doses is consistent with the progressive hydration of surface $K^+$ ions.

**Exposure to ultraclean liquid water**

To bridge the gap between UHV and environmental conditions where minerals are in contact with liquid water, the UHV-cleaved surface of mica was exposed to ultraclean liquid water using the apparatus described briefly in the Experimental section and in detail in Ref. 42. Afterwards, each sample was brought to the main UHV chamber and analyzed either with XPS at room temperature or with nc-AFM at 4.7 K. The surface is clean in XPS except for minor traces of carbon (see inset of Fig. 4c; note that the binding energy of this weak C 1s peak, 286.4 eV, differs significantly from the value of 289.0 eV expected for $K_2CO_3$[43]). The relative intensities of the main core-level peaks are comparable to those of the UHV-cleaved surface apart from the K peaks, which clearly decrease. Taking the K 2p signal of the cleaved surface as a reference and using the same fits for the bulk and surface components, one can estimate that the liquid-water exposure reduces the surface contribution of the K 2p peak to about 63% of its original value. In large-area AFM images (Fig. 4a), the surface appears covered by several clusters of different heights, one of which is marked by an arrow. Atomic resolution is achieved only in small areas in between the protruding clusters. The black square in Fig. 4a marks such an area, exemplified by Fig. 4b. Round, dark features sit on the same hexagonal grid as on the UHV-cleaved mica. However, pronounced contrast differences are appreciable among the dots, and their overall coverage is substantially smaller than in UHV-cleaved mica (between 50% and 80% of the original value, depending on which of the fainter species are considered for the evaluation). The same behavior is observed after a water exposure for ca. 10 minutes instead of one minute.

A tentative interpretation is that some $K^+$ ions become mobile upon exposure to liquid water at room temperature, leaving their aluminosilicate tetrahedra rings (see Fig. 1c). To keep the system charge-neutral, the empty rings become occupied by protons – possibly the fainter species visible in AFM. Because the water dries on the sample, the displaced $K^+$ ions are not rinsed away. The clusters visible in AFM could consist of the displaced $K^+$ ions bound by water molecules (and possibly traces of carbon). These three-dimensional clusters should contribute only marginally to the XPS signal, explaining the measured decrease in the surface contribution of the K 2p peak. Another possibility is that K ions diffuse into mesoscopic cracks into the sample or to the outer rim of the sample following a "coffee-stain effect" when the water drop dries.

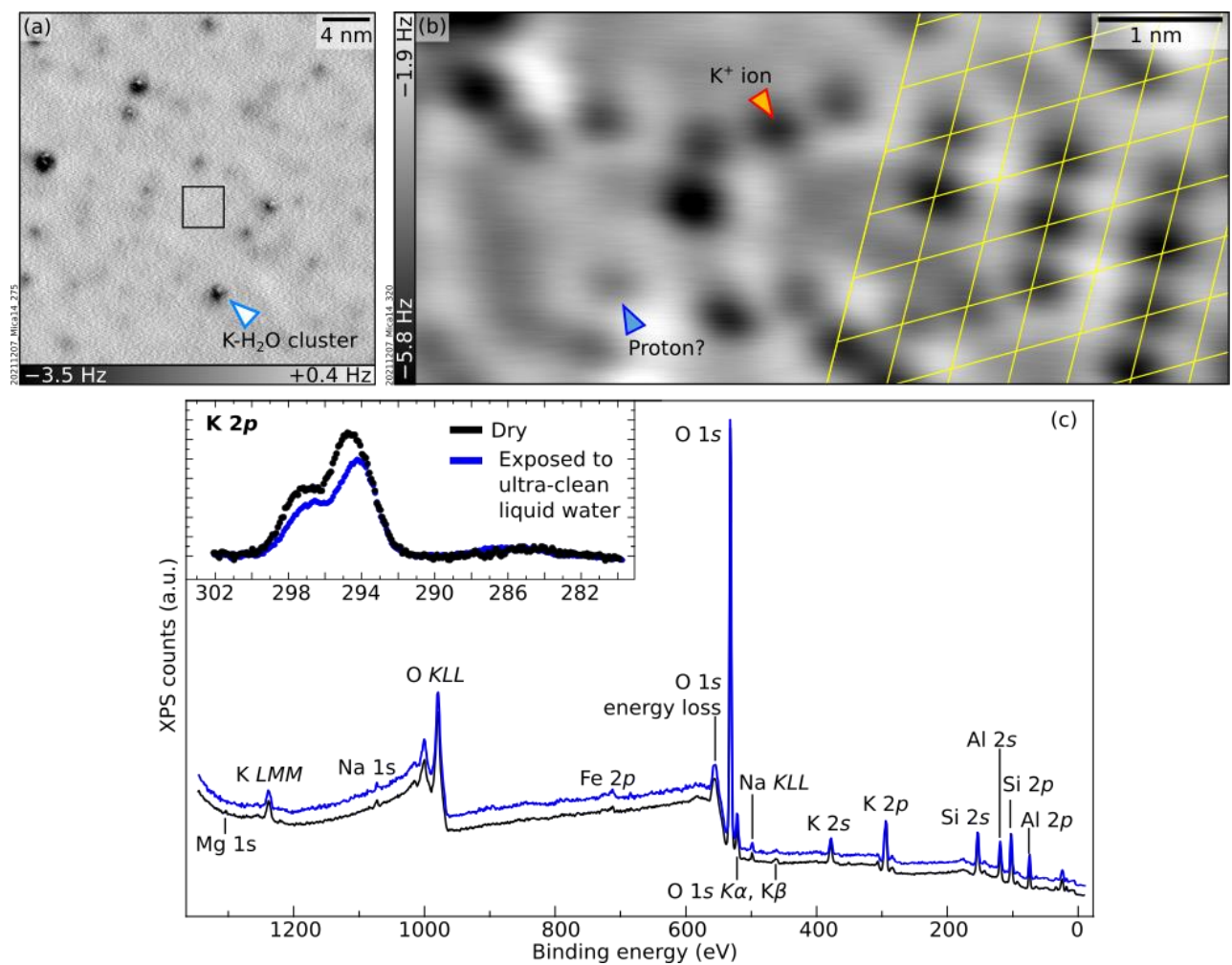

**Figure 4. Mica after exposure to clean liquid water.** (a, b) Constant-height non-contact AFM images. (a) 30 × 30 nm$^2$, $A$ = 300 pm, $V_s$ = −1.2 V; (b) 6.7 × 3 nm$^2$, $A$ = 100 pm, $V_s$ = −1.2 V. (c) XPS survey and K 2$p$ + C 1$s$ region (inset) measured in UHV at room temperature of an as-cleaved mica surface, before (black) and after (blue) exposure to liquid water (Al K$\alpha$, 70° grazing emission, pass energy of 60 eV and 20 eV for the overview and the inset, respectively). In the inset, the Al K$\alpha_3$ contribution of the K 2$p$ signal has been subtracted. The energy axes were adjusted to account for charging (see Experimental section).

The presented data show that experiments performed under ultraclean conditions result only in a partial substitution of the K$^+$ ions after evaporation of the water. This agrees with previous MD simulations which "do not rule out the possibility that the surface K$^+$ ions remain on the surface after washing".[18] In contrast, previous studies where mica was cleaved in air and then rinsed in water reported the complete substitution of the surface K$^+$ ions.[13] The different extents of the ion substitution in UHV vs. ambient conditions could be due to the formation of K$_2$CO$_3$ in ambient conditions. There is evidence that K$_2$CO$_3$ forms on air-cleaved mica due to the reaction of K$^+$ ions with water and atmospheric CO$_2$.[37] K$_2$CO$_3$ is water-soluble and thus likely washed away by rinsing, removing all K from the surface, and leaving no choice to the otherwise charged surface to adsorb protons from the solution.

## Conclusions

This study investigates the interaction of water with the surface K$^+$ ions of cleaved mica under pristine conditions. Hydrated ions are directly visualized with nc-AFM upon dosing water vapor in UHV at 100 K. They appear to be lifted up by water molecules, which orient themselves differently depending on the specific adsorption site. Increasing water coverage causes the clustering of the hydrated species without any apparent ordered ice-like phase. After exposing UHV-cleaved mica to ultraclean water and evaporation of the water, the K$^+$ ions appear to be partially substituted by protons, in contrast with the complete substitution observed in after exposure to liquid water under ambient conditions. The difference could be due to the mediation by K$_2$CO$_3$ formed in ambient conditions.

# Experimental methods and data analysis

## Experimental setup

The experiments were carried out in a UHV setup consisting of three interconnected chambers: a preparation chamber for sample cleaving, water vapor dosing, and XPS measurements (base pressure < $1 \times 10^{-10}$ mbar), an AFM chamber for nc-AFM measurements (base pressure < $2 \times 10^{-11}$ mbar), and a custom-built compartment (hereafter, "side chamber") used for the liquid-drop experiments. The latter is attached to the preparation chamber and has a base pressure < $1 \times 10^{-9}$ mbar after bakeout.

Natural muscovite mica single crystals [(0001)-oriented disks of grade V1, with 10 mm diameter and 0.25 mm thickness, from TedPella] were glued on Omicron-style stainless-steel sample plates with epoxy glue. They were cleaved in UHV before each experiment as described elsewhere.[10]

## Water-vapor experiments

Water vapor was dosed by cooling the sample holder on the manipulator to 100 K with liquid nitrogen. Nominal doses are expressed in Langmuir (L), where 1 L corresponds to a dose obtained by an exposure of 1 s at $1.3 \times 10^{-6}$ mbar (or 100 s at $1.3 \times 10^{-8}$ mbar, the pressure used for the experiments). Determining effective doses on the sample surface is challenging due to (i) the lack of a 2D adsorption pattern and of a theoretical model for the adsorbed phases, and (ii) sticking of $H_2O$ at the chamber walls that leads to an inhomogeneous pressure distribution in the vacuum chamber. Previous calibrations in the same setup on $Fe_2O_3$[44] and $Fe_3O_4$[45] single crystals quantified 1 L dosed at 150 K as 0.8–0.9 $H_2O$ molecules/nm$^2$. Using the 0.234 nm$^2$ unit cell of mica, this gives ≈0.19 $H_2O$/u.c. (or ≈0.38 $H_2O$/surface $K^+$ ion as there are 0.48 K ions per unit cell[10]).

## XPS acquisition and analysis

XPS was acquired with a non-monochromatic dual-anode Mg/Al X-ray source (SPECS XR 50) and a hemispherical analyzer (SPECS Phoibos 100). Spectra were acquired in normal and grazing emission (70° from the surface normal). The insulating nature of the samples (7.85 eV bandgap)[46] makes the XPS analysis challenging, both because of binding energy shifts and broadening effects. After cleaving, the spectra shift to higher binding energies (between 5 and 7 eV); the precise shifts are determined by the amount and type of surface contamination, XPS acquisition geometry, and sample thickness.[47] In this work, dosing water at 100 K caused progressive shifts to lower binding energies. To correct for the charging, the binding energy axes were calibrated such that the position of the K $2p_{3/2}$ component of the bulk corresponded to the literature value of 293.75 eV.[47]

The intensities and positions of the Al-Kα-excited XPS peaks were evaluated with CasaXPS after subtracting a Shirley-type background. Table 1 summarizes the constraints applied for the fits. In detail, the O 1$s$ peaks (Fig. 3a) were fit with two main components: a component obtained by the sum of peaks 1 and 2, fitting the dry surface, and a component (3) fitting the surface covered with 50 L $H_2O$, where the signal from the bulk is negligible (see the corresponding K 2$p$ spectrum) and all water should be molecular. The peak shapes, FWHM values, and relative positions of components 1+2 and 3 were determined from the spectra of the dry surface and of the 50 L-dosed surface, respectively, and were then constrained to fit the other experiments. Peaks 1 and 2 on the dry surface were used to reproduce the asymmetric shape of the corresponding O 1$s$ peak (possibly due to the presence of OH in the subsurface octahedral layer). The position, relative separation, and relative area of components 1 and 2 were constrained to fit the subsequent experiments. One must point out that fitting the O 1$s$ peak poses intrinsic challenges: On the pristine surface, XPS should be sensitive to the OH groups of the subsurface octahedral layer but broadening effects due to charging do not allow to resolve such an OH component. Broadening is likely enhanced on the water-dosed surface. As this work argues, each surface site is characterized by different adsorption geometries of the water molecules; this will cause different screening effects locally, each corresponding to a slightly shifted peak. The measurements average over all possible configurations. Thus, they produce comparatively broader peaks.

The K 2$p$ peaks were fit by multiple sets of 2$p$ peaks. In each set, the separation between 2$p_{3/2}$ and 2$p_{1/2}$ was set to 2.8 eV in line with previous works,[47] and the area ratio to 2:1. All peaks have the same FWHM. The separation between sets (1) and (2) (bulk and surface components, respectively)[10] was set the same for normal and grazing emission.

Table 1. Details about the XPS fitting components of Fig. 3. The shape, full-width half maximum, and position were constrained for all peaks.

|  | Identifier | Shape | FWHM | Position (eV) | Area |
| --- | --- | --- | --- | --- | --- |
| O 1$s$ 1 | O 1$s$ pristine | LA(1.25,243) | 2.24 | 532.28 | Free |
| O 1$s$ 2 | OH pristine | LA(1.25,243) | 1.88 | (O 1$s$ 1) + 0.765 | (Area O 1$s$ 1) × 0.134 |
| O 1$s$ 3 | $H_2O$ | LA(1.25,213) | 1.92 | 533.67 | Free |
| K 2$p$ 1 | K 2$p$ bulk | LA(1,643) | 1.71 | 293.75 (2$p_{3/2}$) | Free |
| K 2$p$ 2 | K 2$p$ surface | LA(1,643) | 1.71 | 295.015 (2$p_{3/2}$) | Free |
| K 2$p$ 3 | K 2$p$ medium coverage | LA(1,643) | 1.71 | 294.5 (2$p_{3/2}$) | Free |

## AFM measurements

The AFM measurements were performed at 4.7 K using a commercial Omicron qPlus LT head and a differential cryogenic amplifier.[48] The tuning-fork-based AFM sensors ($k$ = 3750 N/m, $f_0$ = 45 kHz, Q ≈ 50000)[49] have a separate wire for the tunneling current attached to electrochemically etched W tips, which were cleaned *in situ* by field emission.[50] Before each measurement, the tips were further prepared on a clean Cu(110) single crystal by repeated indentation and voltage pulses. CO-functionalized tips[51] were used to image the water-exposed samples. After a coarse approach, the tip was approached with the AFM frequency controller, setting a typical frequency shift value of −1 Hz. The controller was switched off, and the tip was approached manually until an AFM contrast was visible. All AFM images were acquired in the constant-height mode. At times, the absolute values of frequency shifts obtained during the acquisition of atomically resolved images were large (up to 100 Hz) and not reproducible from sample to sample or on different regions on the same sample. This is because cleaving can result in domains of trapped charges in insulators,[52–54] which can cause long-range electrostatic interactions between the surface and tip.[33] The electrostatic fields can be partially compensated by applying a bias voltage between tip and sample. Most of the measurements were performed at a bias voltage (specified in the respective figure captions) such that the surface was measured as closest as possible to the local contact potential difference between tip and sample, i.e., at the minimum absolute value of the frequency shift.

**Liquid-water experiments**

Samples were exposed to liquid water in the side chamber as detailed elsewhere.[42] In short, the side chamber is equipped with two water reservoirs that can be cooled to 100 K and are connected to the side chamber through all-metal CF16 angle valves. One reservoir is used to make the water drop, the other as a cryosorption pump to remove the water after the experiment. Each experiment proceeds as follows:

- The milliQ water for the drop is purified by three freeze–pump–thaw cycles. The water supply is partially frozen and kept near 0 °C throughout the experiment to stabilize the vapor pressure at 6 mbar evolving from the water-ice mixture.
- After separating the side chamber with a valve, water vapor is dosed for a fixed time inside the side chamber (typically 7 minutes). The water condenses onto a conical, stainless-steel tip held at 100 K where it forms an icicle.
- The as-cleaved mica sample is transferred under UHV into the side chamber while keeping the tip at 100 K. (The vapor pressure of $H_2O$ is negligible at this temperature.)
- The side chamber is closed off again. The tip is heated to room temperature to let the icicle thaw and a liquid droplet fall onto the sample surface. This results in the complete wetting of the sample surface but no water contact with the sample plate (crucial to avoid contamination).
- After exposure, the liquid and residual water vapor in the side chamber are evacuated using the cryosorption pump.
- The sample is transferred back to the preparation chamber. To make sure that most of the residual water evaporates, the sample is kept at room temperature in the preparation chamber for 20 min (or until the pressure in the chamber drops below $4 \times 10^{-10}$ mbar). Afterwards, the sample is transferred to the AFM head. Note that the water drop always dries on the sample's surface, it is not rinsed off like in ex-situ experiments. This is because the strongly hydrophilic nature of UHV-cleaved mica prevents the water drop from coming off the sample even when tilted by 90°.

To minimize the contamination displaced from the chamber walls, the side chamber is baked before use. Before the experiment, to further clean the walls of the side chamber and the tip, a few water drops are created, dropped onto the bottom of the side chamber and pumped away.

## Author Contributions

G. F.: Conceptualization, investigation, writing. S. B.: Investigation. J. B., I. S., J. P.: Investigation, supervision. M. Setvin: Supervision, validation. M. Schmid: Supervision, validation. UD: Funding acquisition, supervision, validation.

## Conflicts of interest

There are no conflicts to declare.

## Acknowledgements

This work was supported by the European Research Council (ERC) under the European Union's Horizon 2020 research and innovation programme (grant agreement No. 883395, Advanced Research Grant 'WatFun'). S.B. acknowledges support from the FFG Project No. 23017619 'FunPakTrio'. M. Setvín was supported through the FWF project "Super" and the Czech Science Foundation, project GACR 20-21727X.

## Notes and references

# Supplementary Material –
# Interaction of surface cations of cleaved mica with water in vapor and liquid forms

*Giada Franceschi, Sebastian Brandstetter, Jan Balajka, Igor Sokolović, Jiri Pavelec, Martin Setvín, Michael Schmid, and Ulrike Diebold*

**S1. Water adsorption at room temperature**

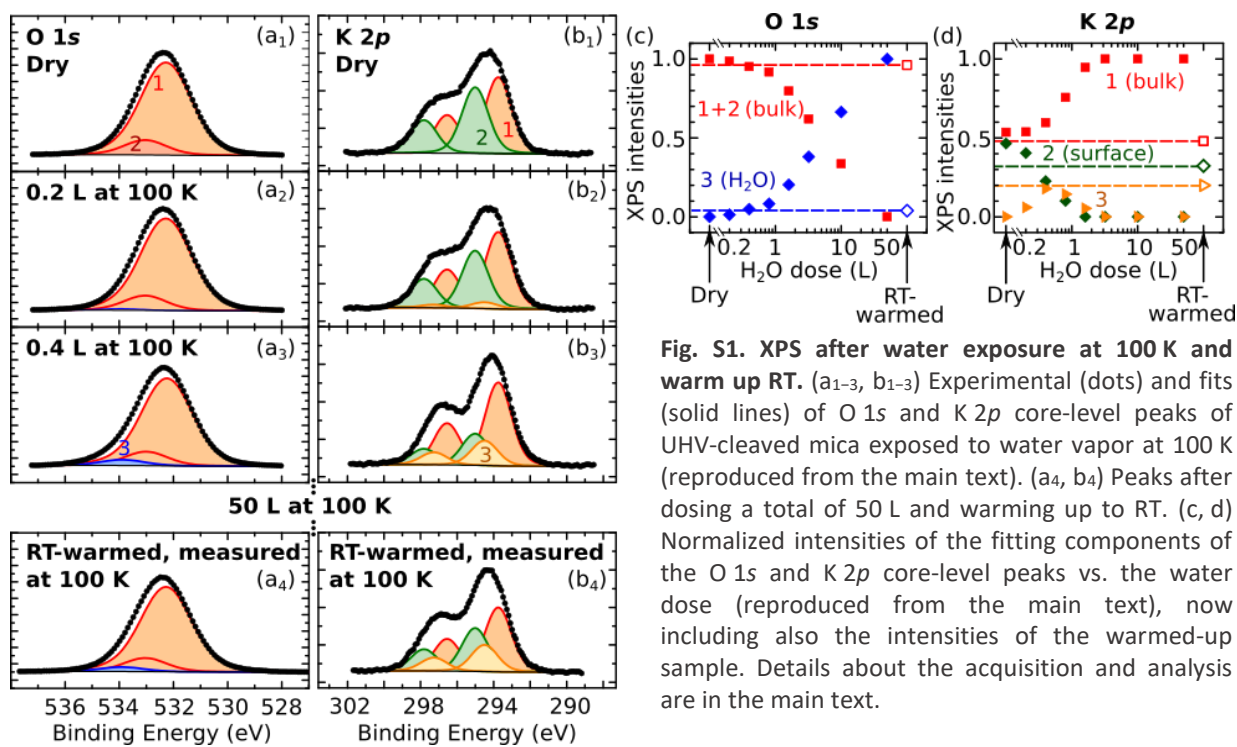

**Fig. S1. XPS after water exposure at 100 K and warm up RT.** ($a_{1-3}$, $b_{1-3}$) Experimental (dots) and fits (solid lines) of O $1s$ and K $2p$ core-level peaks of UHV-cleaved mica exposed to water vapor at 100 K (reproduced from the main text). ($a_4$, $b_4$) Peaks after dosing a total of 50 L and warming up to RT. (c, d) Normalized intensities of the fitting components of the O $1s$ and K $2p$ core-level peaks vs. the water dose (reproduced from the main text), now including also the intensities of the warmed-up sample. Details about the acquisition and analysis are in the main text.

Water does not adsorb at room temperature (RT) on the UHV-cleaved mica. Lowering the sample temperature increases the sticking. However, some water species remain after dosing water vapor at low temperature and warming up to RT, suggesting that the adsorption of water at room temperature is kinetically limited. Figure S1$a_4$,$b_4$ show the K $2p$ and O $1s$ peaks of a sample exposed to 50 L $H_2O$ at 100 K, then warmed up to RT to let the water desorb, and cooled down again for acquiring XPS. (Figs. S1$a_{1-3}$,$b_{1-3}$ correspond to small coverages of water dosed at 100 K; they are reproduced from Fig. 3 of the main text). The peak shape is different from the one of the dry mica, an indication that some water species are still present on the surface (this is consistent with the appearance of nc-AFM images, which do not resemble the as-cleaved surface; not shown). Figures S1c,d include the quantitative analysis of the fitting components of the spectra of the warmed up sample (empty symbols). The values correspond to the ones that can be obtained by dosing between 0.2 L and 0.4 L water at 100 K on UHV-cleaved samples. This indicates that residual water is present on the sample's surface after warming up to RT. Note that acquiring XPS data at room temperature induces a change in the peak shape towards the "dry" spectrum – a sign of X-ray-stimulated desorption occurring at room temperature, possibly due to X-ray-induced electronic transitions.[1] It cannot be excluded that the residual water after warming up to RT is dissociated. Thermally activated dissociation was previously observed on $Fe_3O_4$(001) at around 150 K.[2]

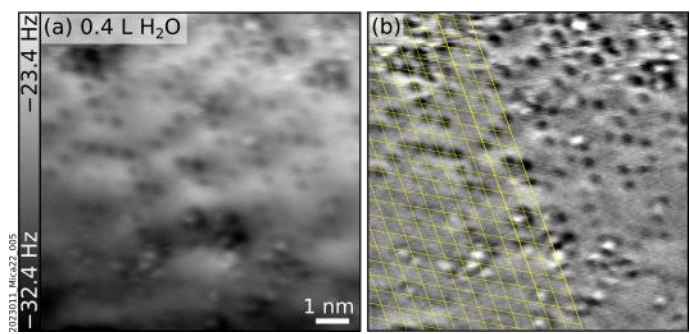

**Fig. S2.** (a, b) 9.6 × 9.6 nm$^2$ nc-AFM image of 0.4 L $H_2O$ dosed at 100 K on a UHV-cleaved mica surface, acquired with an ill-defined tip. (b) Corresponding high-pass-filtered image; the yellow overlay highlights the unit cells of cleaved mica.

**S2. Unusual tip conditions to measure surfaces exposed to water vapor at 100 K**

Figures S2a shows a nc-AFM image of the surface of a UHV-cleaved mica sample after exposure to 0.4 L $H_2O$ at 100 K (the same coverage as Fig. 4b in the main text) acquired with an ill-defined tip – i.e., unintentionally modified after interaction with the surface. Similarly to the surface of dry mica,[3] one can see strong background modulations (this sample was measured at a bias voltage not compensating the LCPD) and resolve black (attractive) dots corresponding to the $K^+$ lattice of cleaved mica. In addition, bright (repulsive) features are visible in the lattice positions "between" the ion sites. The attractive and repulsive features are more evident in the high-pass-filtered image of Figure S2b.

Following the reasonings of the main text, the dark dots are assigned to $K^+$ ions unaffected by the small dose of water vapor introduced in the chamber. With standard tip preparations (Cu, CO-terminated), it was not possible to obtain atomic resolution on the $K^+$ lattice after dosing water vapor on the cleaved surface.

The authors refrain from further analysis of these images since the contrast obtained with this tip was rare.